\documentclass[alpha-refs]{wiley-article}
\usepackage{graphicx}
\usepackage{doi}
\usepackage{booktabs}
\usepackage{array}
\usepackage{calc}


\makeatletter\renewcommand{\footnoterule}{\noindent\textcolor{black!50}{\rule{\columnwidth}{0.5\p@}}\vskip2\p@}\makeatother

\makeatletter
\RenewEnviron{abstract}{\global\let\savedabstract\BODY}
\newcommand{\frontpage}[1]{%
  \maketitle
  \noindent
  \begin{minipage}[t]{\dimexpr 54.9mm-6.5\p@-5mm\relax}\vspace{0pt}%
    \fontsize{7\p@}{10\p@}\selectfont\raggedright\wiley@affilmetadata
  \end{minipage}\hfill
  \begin{minipage}[t]{\dimexpr\textwidth-54.9mm+6.5\p@\relax}\vspace{0pt}%
    \begin{mdframed}[font=\fontsize{9\p@}{15\p@}\selectfont,
      topline=false,bottomline=false,leftline=true,rightline=false,
      backgroundcolor=black!10,middlelinewidth=6\p@,middlelinecolor=white,
      outerlinewidth=0.5\p@]%
      \setlength{\parindent}{1em}\savedabstract
    \end{mdframed}%
    \vskip6\p@
    \noindent{\fontsize{8\p@}{13\p@}\selectfont\textbf{Keywords} -- #1\par}%
  \end{minipage}%
  \vskip2em}
\makeatother

\newtheorem{proposition}{Proposition}
\newtheorem{lemma}{Lemma}
\newtheorem{corollary}{Corollary}
\newtheorem{remark}{Remark}

\papertype{Original Article}

\author[1,2,3]{Eviatar Bach}
\affil[1]{Department of Meteorology and Department of Mathematics and Statistics, University of Reading, Reading, UK}
\affil[2]{Centre for the Mathematics of Planet Earth, University of Reading, Reading, UK}
\affil[3]{National Centre for Earth Observation, Reading, UK}
\corraddress{Eviatar Bach, Brian Hoskins Building, University of Reading, Earley Gate, Whiteknights Rd., Reading, Berkshire, RG6 6ET}
\corremail{eviatarbach@protonmail.com}

\fundinginfo{This work was partially funded by the AUSPICE project, UKRI2165.}

\title{Analysis increments for model bias diagnosis and correction: Theoretical foundations}

\date{}

\begin{document}

\begin{abstract}
A common method for diagnosing and correcting model bias in weather and climate prediction is based on \emph{analysis increments}: the difference between the analysis and the forecast at each data assimilation cycle. Their time average is used to estimate and correct the bias, and machine learning is increasingly used to map the forecast state to the increments, giving a state-dependent correction. Despite its widespread use, the method lacks theoretical foundations.

We first analyse a Kalman filter whose forecast model has a constant bias. We show what the averaged increment measures, and prove that iterating the correction removes all of the bias seen through the observations; for linear dynamics, no method can identify the rest.

We then extend the analysis to state-dependent biases. For linear dynamics, each increment records the model error over the filter's memory, as seen through the observations. We use this to determine what the increments measure when averaged over a long period, averaged over periods with similar slowly varying conditions, and used without averaging, as in machine-learned corrections.

Numerical experiments confirm the theory for a linear model and suggest that its main conclusions carry over to the nonlinear Lorenz '96 model, including for a correction learned by a neural network.
\end{abstract}

\twocolumn[{\frontpage{data assimilation; model bias; analysis increment; bias correction; machine learning; Kalman filter; ensemble Kalman filter}}]
\section{Introduction}

Data assimilation (DA) combines model forecasts with observations to
produce state estimates of dynamical systems \citep{kalnay_atmospheric_2002,sanz-alonso_inverse_2023}. A central assumption of most DA methods is that model errors, if present, have zero mean and are uncorrelated in time. The Kalman filter is optimal under this assumption \citep{kalman_new_1960}, and ensemble Kalman filters are built on it \citep{evensen_ensemble_2003}. In practice, however, forecast models are invariably imperfect, and their errors often have a systematic component, a \emph{model bias}, which in general depends on the state of the system \citep{griffith_adjoint_2000,dee_bias_2005}.

The presence of model bias degrades the quality of the analysis and can
cause filters to diverge in severe cases. Considerable effort has been
devoted to detecting and correcting model bias in DA systems
\citep{friedland_treatment_1969, dee_data_1998, dee_bias_2005}. A key diagnostic tool is the
\emph{analysis increment}: the difference between the
analysis and the forecast at each cycle. Persistent nonzero mean increments
are a clear symptom of model or observation bias \citep{dee_bias_2005},
and time-averaged increments are a standard diagnostic of model error in operational numerical weather prediction.

\begin{figure*}
    \centering
    \includegraphics[width=0.7\linewidth]{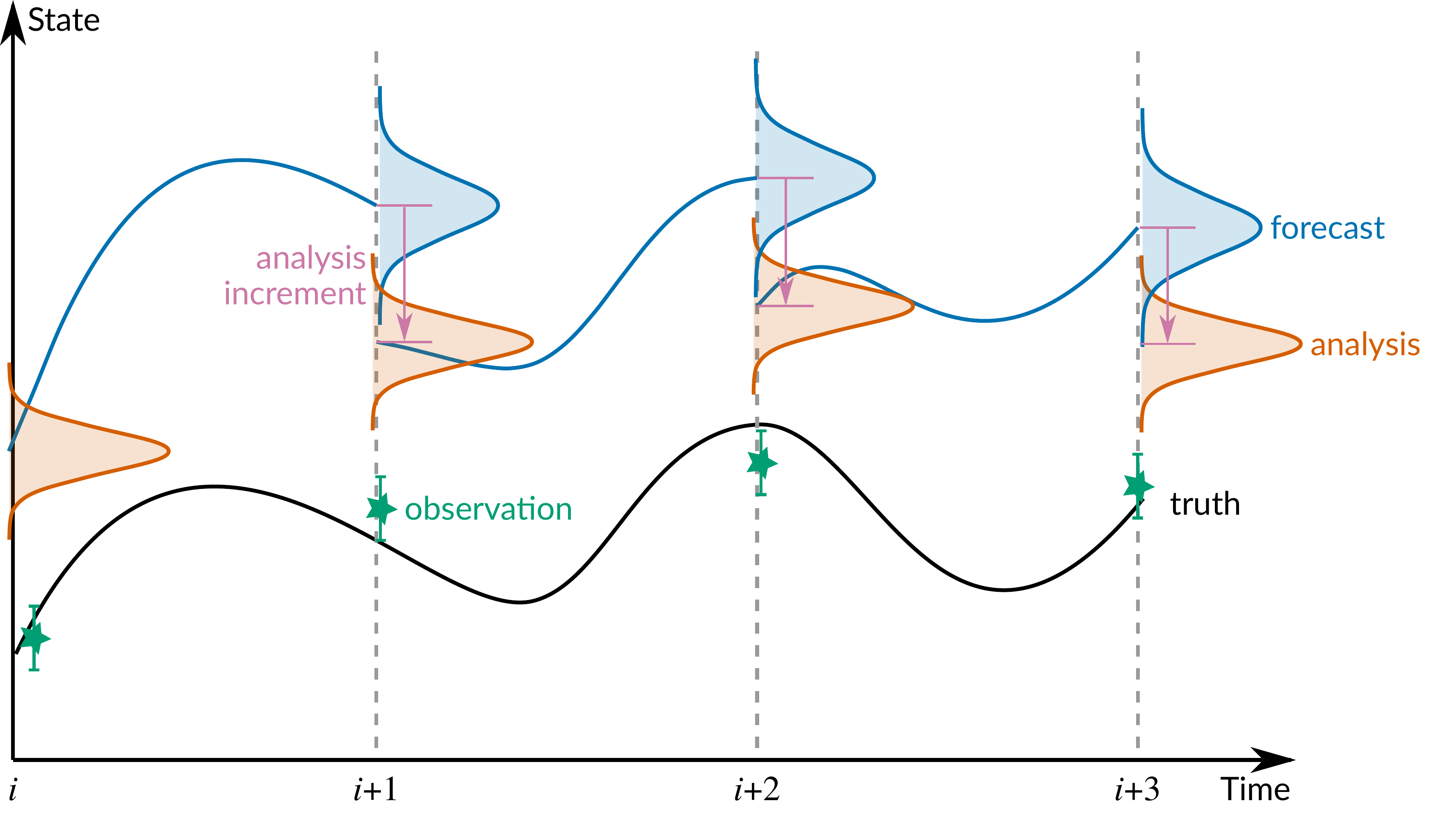}
    \caption{A schematic showing analysis increments in the presence of a systematic model bias. In this case, the forecast model is biased upwards, and the analysis increments point in the opposite direction. Note that the analysis still retains some of the bias. Error bars on the observations show the observation error standard deviation.}
    \label{fig:analysis_increments}
\end{figure*}
The idea behind using analysis increments as a diagnostic tool is the following. If the observations are unbiased, the analysis pulls the forecast towards them, so the analysis increment should point against the model bias; see Figure~\ref{fig:analysis_increments} for an illustration. 

Once diagnosed, the estimated bias can be subtracted from the forecast to correct it. \citet{saha_response_1992} corrected model bias by adding the mean one-day forecast error, relative to the verifying analyses, as a constant forcing in an operational forecast model. \citet{dee_data_1998,dee_bias_2005} analysed forecast bias in cycled data assimilation mathematically, and proposed estimating the bias adaptively during the assimilation. \citet{danforth_estimating_2007} estimated the mean, periodic, and linear state-dependent components of model error from forecast errors relative to a reanalysis, and used them for correction. \citet{bhargava_estimation_2018} used the analysis increments of an operational system to diagnose its mean and periodic model biases. \citet{li_accounting_2009,mitchell_accounting_2015} considered correcting the bias estimated from analysis increments in conjunction with accounting for residual random errors, through additive noise or an estimated model error covariance, and \citet{carrassiTreatmentErrorDue2011} estimated the model error covariance itself from analysis increments.

More recently, machine learning models have been trained to predict analysis increments from the model state, giving a nonlinear, state-dependent correction to the forecast model \citep{bonavita_machine_2020,chen_correcting_2022,amemiya_application_2023,gregory_deep_2023,farchi_development_2025,farchi_batched_2026}. More broadly, the intersection of DA and machine learning is a rapidly growing field, reviewed by \citet{bach_machine_2025, bach_interface_2026, arcucci_convergence_2026}; \citet{levine_framework_2022} give a general framework for learning model error from data.

The popularity of the analysis increment method can be explained by the fact that it is an \emph{offline} one: the bias (whether constant or state-dependent) can be estimated from a time series of analysis increments of an existing assimilation system, such as an operational archive or a reanalysis, without modifying or rerunning that system. \emph{Online} alternatives instead estimate the bias during the assimilation, which requires changing the assimilation system and, for some methods, running it repeatedly; we relate these methods to the increment method in Sections~\ref{sec:unidentifiable} and~\ref{sec:dee}. The distinction matters particularly for learned corrections. Offline, the machine learning model can be trained by ordinary regression on a fixed data set of states and increments; online, training it within the assimilation requires either differentiating through the assimilation \citep{chen_autodifferentiable_2022} or estimating its parameters jointly with the state \citep{gottwald_combining_2021}, which precludes standard gradient-based training and is difficult for models with many parameters.

Despite the widespread practical use of analysis increments as a bias diagnostic and correction method, what the increments measure, and what corrections based on them achieve, have not been established mathematically. Even for a constant bias, the relationship between the time-averaged analysis increment and the asymptotic forecast bias does not appear to have been stated explicitly in the data assimilation literature. How state-dependent biases appear in the increments, and how repeated corrections behave, do not appear to have been analysed either. The method is often justified by the analyses being a proxy for the true state \citep[e.g.,][]{delsole_empirical_1999, danforth_estimating_2007, rodwell_using_2007}, \emph{despite their being contaminated by the bias themselves}. \citet{dee_bias_2005} derives how a forecast bias is transferred to the analysis in a single analysis step, and \citet[][Section~4d]{dee_data_1998} give the recursion for the forecast bias over many cycles of a filter that ignores the bias, but do not relate it to the averaged increment. In the control literature, \citet{lee_discrete-time_1979} derived the steady-state mean error of a Kalman filter with a mismatched model, but did not use it to interpret the filter's increments. The present work fills these gaps.

Our main contributions are:
\begin{enumerate}
    \item For a constant additive bias, we derive the fixed point of the
    asymptotic forecast bias (Proposition~\ref{prop:fixed_point}), and show that
    the time-averaged analysis increment converges to minus the Kalman gain applied
    to the innovation bias \eqref{eq:increment}.
    \item We prove that correcting the forecast model with the averaged increment
    reduces the innovation bias in the norm defined by the innovation
    covariance (Proposition~\ref{prop:mahal}), and that iterating the correction
    drives the innovation bias to zero geometrically
    (Proposition~\ref{prop:iterated_corr}).
    \item We show that the iterated correction removes all of the bias that is
    identifiable from the observations: what remains is equivalent to a constant
    offset of the state in unobserved directions, which, for linear dynamics, no
    method can detect from the observations alone, including augmented-state
    filters (Corollary~\ref{cor:iter_limit}).
    \item For state-dependent biases, we show what the increments measure at
    three levels of averaging. For the averaged increment, we give the exact
    fixed point for linear biases, and
    approximate fixed points for nonlinear biases. We then study how increments averaged by a slowly varying state diagnose the
    slowly varying part of the model error. Finally, we consider individual increments, and what learned corrections regressed on them capture.
    \item We test the theory numerically with a Kalman filter and an ensemble Kalman filter, applied to both linear and nonlinear models, and constant and state-dependent biases.
\end{enumerate}

The paper is organised as follows. Section~\ref{sec:constant_bias} analyses the
constant bias case, which admits a complete analytical treatment and provides
the foundation and baseline for what follows. Section~\ref{sec:statedep} extends
the analysis to state-dependent biases, considering in turn the averaged
increments, increments averaged by a slowly varying state, and learned corrections on individual
increments.
Section~\ref{sec:experiments} presents numerical experiments.
Section~\ref{sec:conclusions} concludes.

\section{Constant bias}\label{sec:constant_bias}
\subsection{Setup}\label{sec:setup}

We follow the notation of \citet{ide_unified_1997} throughout, in which
superscripts $t$, $f$, and $a$ denote the true, forecast, and analysis
states respectively.

We consider a linear state-space model. The true state evolves as
\begin{equation}
    x^\mathrm{t}_{i+1} = M x^\mathrm{t}_i + \eta_i,
    \label{eq:truth}
\end{equation}
where $M \in \mathbb{R}^{n \times n}$ is the model and $\eta_i \sim \mathcal{N}(0, Q)$ is zero-mean
model noise. Observations are given by
\begin{equation}
    y_i = H x^\mathrm{t}_i + \varepsilon_i, \qquad \varepsilon_i \sim \mathcal{N}(0, R),
    \label{eq:obs}
\end{equation}
with $H \in \mathbb{R}^{p \times n}$ and $R$ positive definite. Throughout this
paper we assume $H$ is time-invariant.

The state is estimated with a Kalman filter at steady state. Its forecast model is
imperfect, introducing a constant bias $b \in \mathbb{R}^n$ at each forecast step, which
represents systematic model error. The forecast and analysis steps are
\begin{subequations}
\begin{align}
    x^\mathrm{f}_{i+1} &= M x^\mathrm{a}_i + b, \label{eq:filter}\\
    x^\mathrm{a}_i &= x^\mathrm{f}_i + K(y_i - H x^\mathrm{f}_i),\label{eq:analysis}
\end{align}
\end{subequations}
where $x^\mathrm{f}_i$ and $x^\mathrm{a}_i$ are the means of the filter's forecast
and analysis distributions. The Kalman gain is
\begin{equation*}
    K := P^\mathrm{f} H^T (H P^\mathrm{f} H^T + R)^{-1},
\end{equation*}
where $P^\mathrm{f}$ is the steady-state forecast error covariance.

The filter is unaware of the bias. Its forecast error covariance satisfies the
Riccati equation \eqref{eq:dare} below, which does not involve $b$, so $P^\mathrm{f}$
and $K$ are those of the filter with a perfect model. The analysis error covariance
is $P^\mathrm{a} := (I_n - KH)P^\mathrm{f}$, where $I_n$ is the $n \times n$ identity matrix, and
\begin{align}
    P^\mathrm{f} &= MP^\mathrm{a}M^T + Q \nonumber\\
    &= M\bigl(P^\mathrm{f} - P^\mathrm{f}H^T(HP^\mathrm{f}H^T + R)^{-1}HP^\mathrm{f}\bigr)M^T + Q.
    \label{eq:dare}
\end{align}

Throughout, we assume that $(M, H)$ is detectable and $(M, Q^{1/2})$ is
stabilisable (the latter holds, for instance, if $Q$ is positive
definite). Then \eqref{eq:dare} has a unique
positive semidefinite solution, the filter's covariance converges to it from any
initial covariance, and the steady-state filter is stable: the matrix
$\Phi := M(I_n - KH)$ satisfies $\rho(\Phi) < 1$, where $\rho(\Phi)$ denotes the spectral radius of $\Phi$
\citep[][Section~4.4]{anderson_optimal_1979}.
Since the gain converges to its steady-state value from any initial covariance,
assuming the filter to be at steady state, as above, does not affect the asymptotic
results that follow. With its gain
constant, the filter is the same as cycled three-dimensional variational
assimilation (3D-Var) with a static background error covariance equal to
$P^\mathrm{f}$ \citep[][Sections 5.4.1 and 5.5.1]{kalnay_atmospheric_2002}. We also
assume that $P^\mathrm{f}$ is positive definite. This holds if $Q$ is positive
definite, since $P^\mathrm{f} = MP^\mathrm{a}M^T + Q$ by \eqref{eq:dare}.

Throughout, $\mathbb{E}$ denotes expectation over realisations of the model noise,
the observation noise, and the initial condition. The filter's estimates
$x^\mathrm{f}_i$ and $x^\mathrm{a}_i$ are computed from the observations and are
therefore random variables in this sense.

\subsection{Asymptotic forecast bias}\label{sec:asymptotic_bias}

Define the forecast bias at step $i$ as $b^\mathrm{f}_i := \mathbb{E}[x^\mathrm{f}_i - x^\mathrm{t}_i]$.
Taking expectations of the analysis step \eqref{eq:analysis} gives the analysis bias:
\begin{equation}\label{eq:analysis_bias}
    b^\mathrm{a}_i := \mathbb{E}[x^\mathrm{a}_i - x^\mathrm{t}_i]
    = b^\mathrm{f}_i - KH b^\mathrm{f}_i = (I_n - KH) b^\mathrm{f}_i.
\end{equation}
This equation was also derived in \citet{dee_data_1998,dee_bias_2005}.
The analysis bias is thus the forecast bias multiplied by $(I_n - KH)$. Propagating through the model to the next forecast:
\begin{equation*}
    b^\mathrm{f}_{i+1} = \mathbb{E}[x^\mathrm{f}_{i+1} - x^\mathrm{t}_{i+1}]
    = M b^\mathrm{a}_i + b = M(I_n - KH) b^\mathrm{f}_i + b.
\end{equation*}
The bias therefore satisfies the linear recurrence relation
\begin{equation}
    b^\mathrm{f}_{i+1} = \Phi\, b^\mathrm{f}_i + b, \qquad \Phi = M(I_n - KH).
    \label{eq:recursion}
\end{equation}

We can now obtain the fixed point of this recurrence relation.

\begin{proposition}[Asymptotic forecast bias]\label{prop:fixed_point}
    Under the assumptions of Section~\ref{sec:setup}, the forecast bias
    converges to the fixed point
    \begin{equation}
        b^* = (I_n - \Phi)^{-1} b = (I_n - M(I_n-KH))^{-1} b,
        \label{eq:fixedpoint}
    \end{equation}
    and the analysis bias converges to $(I_n - KH) b^*$.
\end{proposition}

\begin{proof}
    Since $\rho(\Phi) < 1$, the recurrence relation \eqref{eq:recursion} converges to the unique fixed point $b^*$ satisfying
    $b^* = \Phi b^* + b$, i.e., $(I_n - \Phi) b^* = b$ (see, e.g., Theorem 4.13 in \citet{elaydi_introduction_2005}). Since
    $\rho(\Phi) < 1$ implies $1$ is not an eigenvalue of $\Phi$, the matrix
    $(I_n - \Phi)$ is invertible and $b^* = (I_n - \Phi)^{-1} b$. The
    analysis bias follows from \eqref{eq:analysis_bias}.
\end{proof}

\citet[][Section~4d]{dee_data_1998} give the
recursion \eqref{eq:recursion}, for a time-varying gain, in their eq.~(79). Their
eq.~(87), the condition on the per-step bias under which the forecast bias is
constant, is \eqref{eq:fixedpoint} rearranged, although they do not use it to
obtain the asymptotic forecast bias or show that the forecast bias converges to
it. \citet[][eq.~(29)]{lee_discrete-time_1979} derive an
equivalent expression for the steady-state mean error of a mismatched
steady-state Kalman filter, assuming that both the true and the model dynamics
are stable; here the assumptions of Section~\ref{sec:setup} suffice, and $M$ may
be unstable.

\subsection{The time-averaged analysis increment}\label{sec:increment}

The analysis increment at step $i$ is $x^\mathrm{a}_i - x^\mathrm{f}_i = K d_i$,
where $d_i := y_i - Hx^\mathrm{f}_i$ is the innovation. Taking expectations:
\begin{equation*}
    \mathbb{E}[x^\mathrm{a}_i - x^\mathrm{f}_i] = K \mathbb{E}[d_i]
    = KH\,\mathbb{E}[x^\mathrm{t}_i - x^\mathrm{f}_i] = -KH b^\mathrm{f}_i.
\end{equation*}
Up to sign, the asymptotic mean of the innovations is therefore $Hb^*$, which we call
the innovation bias.

We now connect the expected analysis increment to the time-averaged one. Define the average of the analysis increments from time 1 to $T$ as
\begin{equation*}
    [\overline{x^\mathrm{a} - x^\mathrm{f}}]_T := \frac{1}{T} \sum_{i=1}^T \bigl(x^\mathrm{a}_i - x^\mathrm{f}_i\bigr),
\end{equation*}
and write $\overline{x^\mathrm{a} - x^\mathrm{f}} := \lim_{T\to\infty}[\overline{x^\mathrm{a} - x^\mathrm{f}}]_T$.
Write each increment as a deterministic part plus a fluctuation,
\begin{equation}
    \label{eq:increment_decomposition}
    x^\mathrm{a}_i - x^\mathrm{f}_i = -KHb^\mathrm{f}_i + \nu_i, \qquad
    \nu_i := K\bigl(d_i - \mathbb{E}[d_i]\bigr).
\end{equation}
Since the dynamics are linear, the bias only shifts the mean of the innovations. The centred innovations $d_i - \mathbb{E}[d_i]$ are those the steady-state filter would produce with a perfect model, which, at stationarity, are independent and identically distributed as $\mathcal{N}(0,S)$, with $S = HP^\mathrm{f}H^T + R$ \citep[][Section~5.3, Theorem~3.1, and Section~5.4]{anderson_optimal_1979}. The time average of the fluctuations $\nu_i$ therefore tends to zero, by the law of large numbers. Since $b^\mathrm{f}_i \to b^*$, the time average of the deterministic part $-KHb^\mathrm{f}_i$ tends to $-KHb^*$. Hence
\begin{equation}
    \overline{x^\mathrm{a} - x^\mathrm{f}} = -KH b^*.
    \label{eq:increment}
\end{equation}
For a single analysis, \citet[][Section~2]{dee_data_1998} noted that the mean
increment underestimates the forecast bias, since the gain coefficients are
generally less than one. Equation~\eqref{eq:increment} makes this precise for the cycled
filter, in which the forecast bias itself has accumulated to $b^*$.

The averaged increment does not see all of the bias. The matrix
$KH$ has the same null space as $H$: if $KHz = 0$ then,
since $P^\mathrm{f}$ is invertible, $(Hz)^TS^{-1}(Hz) = 0$, so $Hz = 0$. By \eqref{eq:increment},
only the component of $b^*$ in the row space of $H$ is therefore detectable from the
averaged analysis increment. The component in the null space
$\ker(H) = \{z : Hz = 0\}$, that is, in state directions not seen by the
observations, is invisible to it; such directions exist whenever
$\mathrm{rank}(H) < n$, in particular when $p < n$. This does not mean that
unobserved variables are left uncorrected. The averaged increment $-KHb^*$ is
generally nonzero in unobserved variables, because the gain spreads the observed bias into them through the
forecast-error covariances, just as the analysis updates unobserved state
variables.

\begin{remark}[Rate of convergence of the time average]\label{rem:rate}
    Since the fluctuations $\nu_i$ are, at stationarity, independent with covariance $KSK^T$, their
    time average over $T$ cycles has covariance $KSK^T/T$, so the error in
    estimating $-KHb^*$ from $T$ cycles is $O(T^{-1/2})$. In addition, the
    initial transient $b^\mathrm{f}_i - b^* = \Phi^i(b^\mathrm{f}_0 - b^*)$ decays
    geometrically, since $\rho(\Phi) < 1$, and so biases the average by $O(1/T)$;
    the decay is slow when $\rho(\Phi)$ is close to one. For suboptimal
    filters, the innovations are serially
    correlated, and the asymptotic covariance will differ.
\end{remark}

\begin{remark}[Unbiasedness of observations]
A critical assumption of the above analysis is that the observations are
unbiased, i.e., $\mathbb{E}[\varepsilon_i] = 0$. If the observations
have a systematic bias $b^o = \mathbb{E}[\varepsilon_i] \neq 0$, the
innovation satisfies
\begin{equation*}
    \mathbb{E}[d_i] = \mathbb{E}[y_i - Hx^\mathrm{f}_i]
    = H\,\mathbb{E}[x^\mathrm{t}_i - x^\mathrm{f}_i] + b^o = -Hb^\mathrm{f}_i + b^o,
\end{equation*}
and the averaged analysis increment becomes
\begin{equation*}
\overline{x^\mathrm{a} - x^\mathrm{f}} =
-KH b^* + Kb^o,
\end{equation*}
with $b^*$ now the modified fixed point given below,
conflating the forecast bias $b^*$ with the observation
bias $b^o$ projected through $K$. The fixed-point equation
\eqref{eq:fixedpoint} is modified to
\begin{equation}
    b^* = (I_n-\Phi)^{-1}(b + MK b^o),
    \label{eq:obs_bias_fp}
\end{equation}
so that observation bias acts as an additional effective per-step model
bias $MKb^o$, since the analysis bias $(I_n-KH)b^\mathrm{f}_i + Kb^o$ is
propagated by $M$ to the next forecast. In this case the averaged increment
correction does not distinguish between model and observation bias, and
applying it to the forecast corrects the combined effect rather than
the model bias alone. \citet{fowlerRobustnessMethodsAccount2024} studies how observation bias affects the performance of bias correction methods.
\end{remark}

\subsection{Bias correction using the averaged analysis increment}\label{sec:bias_correction}

A natural bias correction scheme is to use the time-averaged analysis increment
as an estimate of the bias. Since the increment points against the bias, the
estimate is, by \eqref{eq:increment},
\begin{equation*}
    \hat{b} := -\overline{x^\mathrm{a} - x^\mathrm{f}} = KHb^*,
\end{equation*}
and the forecast is corrected by subtracting it at each cycle:
\begin{equation}
    \tilde{x}^\mathrm{f}_{i+1} = Mx^\mathrm{a}_i + b - \hat{b}
                       = Mx^\mathrm{a}_i + b - KHb^*.
    \label{eq:corrected_forecast}
\end{equation}
Here $b$ is the actual per-step bias of the forecast model and $\hat{b}$ its
estimate. Using \eqref{eq:fixedpoint}, the effective per-step bias after
correction is:
\begin{equation*}
    \tilde{b} = b - \hat{b} = \bigl(I_n - KH(I_n-\Phi)^{-1}\bigr)b.
\end{equation*}
The new asymptotic bias satisfies $\tilde{b}^* = (I_n-\Phi)^{-1}\tilde{b}$,
giving:
\begin{equation}
    \tilde{b}^* = \bigl(I_n - (I_n-\Phi)^{-1}KH\bigr)b^*.
    \label{eq:corrected_bias}
\end{equation}

Is the asymptotic bias improved after correction? The following proposition shows that the correction does in fact reduce the innovation bias in the Mahalanobis norm with respect to the innovation covariance.

\begin{proposition}[Mahalanobis norm reduction]\label{prop:mahal}
    Under the assumptions of Section~\ref{sec:setup},
    \begin{equation*}
        \|H\tilde{b}^*\|_{S} \leq \|Hb^*\|_{S},
    \end{equation*}
    where $\|v\|_{S}^2 := v^TS^{-1}v$ is the Mahalanobis norm with
    respect to the innovation covariance $S = HP^\mathrm{f}H^T + R$. The
    inequality is strict whenever $Hb^* \neq 0$.
\end{proposition}
\begin{proof}
    In Appendix~\ref{appendix:proof}.
\end{proof}

\subsection{Iterated correction}\label{sec:iterated}

The correction \eqref{eq:corrected_forecast} can be applied iteratively
as follows: at iteration $j$, run the filter to convergence, compute the time-averaged
analysis increment, subtract the corresponding bias estimate from the
already-corrected forecast model, and repeat. Unlike a single correction, this is no
longer offline, since each iteration requires a new assimilation run. To our knowledge, this procedure has only been applied empirically in the climate modelling literature \citep{krinner_iterative_2025}, and, with learned state-dependent corrections, by \citet{gregory_machine_2024} (see Section~\ref{sec:nl_bound}).

At iteration $j$, the filter is run with the forecast model
\begin{equation*}
    x^{\mathrm{f},(j)}_{i+1} = Mx^{\mathrm{a},(j)}_i + b^{(j)}, \qquad b^{(0)} = b,
\end{equation*}
and the per-step bias is updated by subtracting the bias estimate of that run,
$\hat{b}^{(j)} := -\overline{x^{\mathrm{a},(j)} - x^{\mathrm{f},(j)}} = KHb^{*(j)}$:
\begin{equation*}
    b^{(j+1)} = b^{(j)} - \hat{b}^{(j)} = b^{(j)} - KH b^{*(j)},
\end{equation*}
where $b^{*(j)} = (I_n-\Phi)^{-1}b^{(j)}$ is the asymptotic forecast bias of
iteration $j$. Equivalently, $b^{(j)} = b - \sum_{k=0}^{j-1}\hat{b}^{(k)}$: the
forecast model at iteration $j$ subtracts the sum of all previous bias
estimates. The new asymptotic bias satisfies:
\begin{equation}
    b^{*(j+1)} = (I_n-\Phi)^{-1}b^{(j+1)}
    = \bigl(I_n - (I_n-\Phi)^{-1}KH\bigr)b^{*(j)}.
    \label{eq:iter_bias}
\end{equation}
Unrolling the recurrence \eqref{eq:iter_bias} from $b^{*(0)} = b^*$:
\begin{equation*}
    b^{*(j)} = \bigl(I_n - (I_n-\Phi)^{-1}KH\bigr)^j b^*,
\end{equation*}
which does not in general converge to zero. The matrix
$(I_n-\Phi)^{-1}KH$ vanishes on the null space of $H$, so
$I_n - (I_n-\Phi)^{-1}KH$ leaves any $b^* \in \ker(H)$ unchanged: as noted in
Section~\ref{sec:increment}, bias that is invisible in the observations is not removed,
no matter how many iterations are performed.

The innovation bias $Hb^{*(j)}$ obeys a closed recurrence. Since
$H(I_n-\Phi)^{-1}KH = CH$ with
\begin{equation}
    C := H(I_n-\Phi)^{-1}K \in \mathbb{R}^{p\times p},
    \label{eq:C}
\end{equation}
applying $H$ to \eqref{eq:iter_bias} gives $Hb^{*(j+1)} = (I_p - C)\,Hb^{*(j)}$,
and hence
\begin{equation}
    Hb^{*(j)} = (I_p - C)^j\, Hb^*.
    \label{eq:iter_innov}
\end{equation}

The following proposition characterises the convergence of the iterated correction in observation space.

\begin{proposition}[Convergence of the iterated correction]\label{prop:iterated_corr}
    Under the assumptions of Section~\ref{sec:setup}, the iterated correction
    drives the innovation bias to zero geometrically:
    $(I_p - C)^j v \to 0$ for every $v$ of the form $Hz$, and in particular
    $\|Hb^{*(j)}\|_2\to 0$ as $j\to\infty$. If $H$ has full row rank ($\mathrm{rank}(H) = p$, that is, no
    observation is a linear combination of the others), the rate
    is $\rho(I_p - C) < 1$.
\end{proposition}
\begin{proof}
    In Appendix~\ref{appendix:proof}.
\end{proof}

\begin{remark}[Single-step norm reduction and transient growth]
    Proposition~\ref{prop:mahal} does not extend to the Euclidean norm: a single
    correction can increase $\|Hb^*\|_2$. When $H$ has full row rank, a
    single-step reduction for every $b^*$ is equivalent to
    $\sigma_{\max}(I_p-C) \leq 1$, and this fails for some systems in which $C$
    is sufficiently non-normal. The iterates $(I_p-C)^jHb^*$ still converge to zero by
    Proposition~\ref{prop:iterated_corr}, but not necessarily monotonically,
    a phenomenon known as transient growth \citep{trefethen_spectra_2005}.
\end{remark}

Proposition~\ref{prop:iterated_corr} also implies that $C$ is invertible on
vectors of the form $Hz$, and we write $C^{-1}$ for this
inverse.\footnote{By Proposition~\ref{prop:iterated_corr}, $C$ sends no nonzero
vector of the form $Hz$ to zero, since $I_p - C$ would leave such a vector
unchanged, and so would not drive it to zero. Since $C$ maps such vectors to
vectors of the same form, it is invertible on them. More precisely, $C^{-1}$
denotes $\bigl(C|_{\mathrm{range}(H)}\bigr)^{-1}$, the inverse of the restriction
of $C$ to the range of $H$.} It is the usual inverse when $H$ has full row rank.

\begin{corollary}[Limit of the iterated correction]\label{cor:iter_limit}
    Under the assumptions of Section~\ref{sec:setup}, the per-step bias of the
    iterated correction converges to
    \begin{subequations}
    \begin{equation}
        b^{(\infty)} = b - KC^{-1}Hb^*,
    \label{eq:iter_limit}
\end{equation}
    and the asymptotic bias converges to $\Pi b^*$, where
    \begin{equation}
        \Pi := I_n - (I_n-\Phi)^{-1}KC^{-1}H.
    \label{eq:Pi}
\end{equation}
    \end{subequations}
    The limit $\Pi b^*$ is invisible to the observations ($H\Pi b^* = 0$), and the
    remaining per-step bias is of the form $b^{(\infty)} = (I_n-M)z$ with $Hz = 0$.
\end{corollary}
\begin{proof}
    In Appendix~\ref{appendix:proof}.
\end{proof}

\subsection{What cannot be identified}\label{sec:unidentifiable}

What does the remaining bias of Corollary~\ref{cor:iter_limit} mean? Take the true
trajectory and shift every state by a constant vector $z$ with $Hz = 0$, as in
Corollary~\ref{cor:iter_limit}: $x_i := x^\mathrm{t}_i + z$. The shifted sequence
obeys
\begin{align*}
    x_{i+1} = x^\mathrm{t}_{i+1} + z &= Mx^\mathrm{t}_i + \eta_i + z\\
    &= M(x_i - z) + \eta_i + z = Mx_i + (I_n-M)z + \eta_i,
\end{align*}
which is the forecast model with per-step bias $(I_n-M)z$: a model with this bias
reproduces the true trajectory shifted by $z$. Since $Hz = 0$, the shift is
invisible in the observations at every time:
$Hx_i = H(x^\mathrm{t}_i + z) = Hx^\mathrm{t}_i$ for all $i$. We write
$(I_n-M)\ker(H)$ for the set of such per-step biases.

For linear dynamics, no method that uses only the observations can therefore
detect a bias in this set: the models with per-step biases $b$ and
$b + (I_n-M)z$, with $Hz = 0$, give the same distribution of observations, apart
from the effect of the initial condition. In control-theoretic terms, these biases
are undetectable inputs, whose effect on the outputs can be reproduced by a change
of initial state \citep[][Lemma~3.1]{pasqualetti_attack_2013}. The iterated correction
thus removes all of the bias that is identifiable from the observations. Note that
$(I_n-M)\ker(H)$ differs from the set of biases in unobserved variables
($Hb = 0$), which are in general identifiable, since $M$ propagates them into the
observed directions.

Online methods, which estimate the bias during the assimilation, meet the same
limit. In augmented-state filters, which treat the bias as an additional state
variable \citep{friedland_treatment_1969, dee_data_1998}, these directions are
unobservable modes of the augmented system
\citep[][Lemma~1]{pannocchia_disturbance_2003}; and for likelihood-based methods
\citep{dee_maximum-likelihood_1999, tandeo_review_2020, chen_autodifferentiable_2022},
all biases that differ by an element of this set are equally likely given the
observations, apart from a contribution from the prior on the initial state that
does not grow with the length of the record, so maximising the likelihood cannot
determine that part of the bias.

\begin{remark}[Nonlinear dynamics]\label{rem:nonlinear_ident}
    For a nonlinear model $\mathcal{M}$, the shift argument does not apply in
    general: shifting the state by $z$ corresponds to the state-dependent bias
    $z + \mathcal{M}(x) - \mathcal{M}(x+z) \approx (I_n - J(x))z$, with $J(x)$ the
    Jacobian of $\mathcal{M}$ at $x$, which a constant bias can reproduce only if
    $J(x)z$ does not depend on $x$. Bias in these directions then becomes
    identifiable through the variation of $J(x)z$ over the attractor.
\end{remark}

\subsection{Relation to adaptive bias correction}\label{sec:dee}

The iterated correction of Section~\ref{sec:iterated} is closely related to the
adaptive bias correction scheme of \citet[][Section~3b]{dee_bias_2005}, which can
be written as
\begin{subequations}
\begin{align}
    x^\mathrm{f}_{i+1} &= Mx^\mathrm{a}_i + b - \hat{b}_i, \label{eq:dee}\\
    \hat{b}_i &= \hat{b}_{i-1} - \gamma\,(x^\mathrm{a}_i - x^\mathrm{f}_i), \qquad \hat{b}_0 = 0,
\end{align}
\end{subequations}
where the analysis step is \eqref{eq:analysis} as before and $\gamma > 0$ is a parameter that controls how much recent information is
weighted in the running bias estimate $\hat{b}_i$. The forecast model thus has a
per-step bias $b - \hat{b}_i$ that is updated after every cycle by $\gamma$ times the
analysis increment, whereas the iterated correction updates it after a whole run
by the averaged increment: the scheme is an online counterpart of the iterated
correction. If it converges (which requires $\gamma$ to be sufficiently small), it reaches the same limit, $\Pi b^*$, and so likewise
leaves only the unidentifiable component of the bias.

To see this, let $b^\mathrm{f}_i = \mathbb{E}[x^\mathrm{f}_i - x^\mathrm{t}_i]$ as
before and $\beta_i := \mathbb{E}[\hat{b}_i]$ be the mean of the (random) bias
estimate. Taking expectations gives
\begin{equation*}
    b^\mathrm{f}_{i+1} = \Phi b^\mathrm{f}_i + b - \beta_i, \qquad
    \beta_i = \beta_{i-1} + \gamma KHb^\mathrm{f}_i,
\end{equation*}
which is the recurrence \eqref{eq:recursion} with the per-step bias $b$ replaced by
the corrected per-step bias $b - \beta_i$.
The fixed points are those with $KHb^\mathrm{f} = 0$, i.e., $Hb^\mathrm{f} = 0$, since
$KH$ has the same null space as $H$ (Section~\ref{sec:increment}), so that the innovation
bias vanishes, as for the iterated correction. The fixed point is moreover unique:
since $\hat{b}_0 = 0$, the update gives $\beta_i = Kv_i$ with
$v_i = H\bigl(\gamma\sum_{k\le i} b^\mathrm{f}_k\bigr)$ of the form $Hz$. At a fixed
point, $b^\mathrm{f} = (I_n-\Phi)^{-1}(b - Kv)$, and applying $H$ gives
$Hb^\mathrm{f} = Hb^* - Cv$, so $Hb^\mathrm{f} = 0$ requires $Cv = Hb^*$, that is,
$v = C^{-1}Hb^*$, since $C$ is invertible on vectors of the form $Hz$. Hence
$b^\mathrm{f} = \Pi b^*$, as in Corollary~\ref{cor:iter_limit}.

\section{State-dependent bias}
\label{sec:statedep}

When the model error depends on the state, the increments can be used at three
levels of averaging: averaged over the whole record, as in a constant correction (Section~\ref{sec:averaged}); averaged over states with similar
slowly varying components (Section~\ref{sec:slow}); or without averaging, as training
data for a learned state-dependent correction (Section~\ref{sec:nl_bound}). The
first follows from the mean of the filter equations, as for a constant bias; the
other two rest on an exact decomposition of each increment
(Section~\ref{sec:pathwise}).

\subsection{Setup}\label{sec:statedep_setup}

We keep the setup of Section~\ref{sec:setup}. The only change is
that the bias of the forecast model now depends on the state at the start of the
forecast step:
\begin{equation}
    x^\mathrm{f}_{i+1} = Mx^\mathrm{a}_i + b(x^\mathrm{a}_i),
    \label{eq:sd_model}
\end{equation}
where $b : \mathbb{R}^n \to \mathbb{R}^n$ is an unknown function; the constant
bias of Section~\ref{sec:constant_bias} is the case $b(x) \equiv b$.

When relating averages over a long record to expectations, we assume that the
joint process of the truth and the filter converges to a stationary distribution and
is ergodic, so that averages of the
increments over a long record converge to the corresponding expectations as $i \to \infty$.
For the linear truth \eqref{eq:truth}, this requires $\rho(M) < 1$, so the true state
has zero mean at stationarity.

\subsection{Averaged increments}\label{sec:averaged}

Taking expectations of the
analysis step gives, as in Section~\ref{sec:asymptotic_bias}, the analysis bias
$(I_n-KH)b^\mathrm{f}_i$, with
$b^\mathrm{f}_i = \mathbb{E}[x^\mathrm{f}_i - x^\mathrm{t}_i]$, and the expected
increment $-KHb^\mathrm{f}_i$. Since
$x^\mathrm{f}_{i+1} - x^\mathrm{t}_{i+1} = M(x^\mathrm{a}_i - x^\mathrm{t}_i)
+ b(x^\mathrm{a}_i) - \eta_i$, the forecast bias satisfies
\begin{equation}
    b^\mathrm{f}_{i+1} = \Phi b^\mathrm{f}_i + \mathbb{E}[b(x^\mathrm{a}_i)].
    \label{eq:sd_exact}
\end{equation}
Averaging the increments over a long record therefore gives
$-KHb^*$, where $b^* = \lim_{i\to\infty} b^\mathrm{f}_i
= (I_n-\Phi)^{-1}\lim_{i\to\infty}\mathbb{E}[b(x^\mathrm{a}_i)]$. This limit exists by the
assumption of Section~\ref{sec:statedep_setup}, and it is the asymptotic forecast bias, that is, the expected forecast error at stationarity; by ergodicity, it is also the time average of the forecast error over a long record. The averaged increment thus reflects the model error
averaged over the states that the analysis visits, which can differ substantially
from the model error at the mean state. A constant correction based on it therefore
removes only this climatological average; where the model error changes sign over
the visited states, much of it averages out, and only a state-dependent correction
can remove it.

Unlike in the constant bias case, the recurrence \eqref{eq:sd_exact} is not
closed: the term $\mathbb{E}[b(x^\mathrm{a}_i)]$ depends on the full distribution
of $x^\mathrm{a}_i$, not just its mean, which is set by $b^\mathrm{f}_i$. Closing it
requires either linearity of $b$ or an approximation, which we discuss below. The
resulting fixed-point equations for the asymptotic forecast bias $b^*$ are a tool for
understanding what the increments measure, not something to be solved in practice.

\subsubsection{Linear bias}\label{sec:linear_sd}

Suppose $b(x) = Bx + c$ for $B \in \mathbb{R}^{n \times n}$ and $c \in \mathbb{R}^n$.\footnote{Strictly speaking,
this is an affine function of the state; however, we follow the usual terminology and call it a linear bias.}
Then $\mathbb{E}[b(x^\mathrm{a}_i)] = B\mathbb{E}[x^\mathrm{a}_i] + c$, and
since $\mathbb{E}[x^\mathrm{a}_i] = \mathbb{E}[x^\mathrm{t}_i] + b^\mathrm{a}_i
= \mathbb{E}[x^\mathrm{t}_i] + (I_n-KH)b^\mathrm{f}_i$, the recurrence
\eqref{eq:sd_exact} closes:
\begin{equation}
    b^\mathrm{f}_{i+1} = \bigl(\Phi + B(I_n-KH)\bigr)b^\mathrm{f}_i
    + B\,\mathbb{E}[x^\mathrm{t}_i] + c.
    \label{eq:linear_rec}
\end{equation}
If $\rho\bigl(\Phi + B(I_n-KH)\bigr) < 1$, then, since $\mathbb{E}[x^\mathrm{t}_i] \to 0$
(Section~\ref{sec:statedep_setup}), the forecast bias converges to
\begin{equation}
    b^* = \bigl(I_n - \Phi - B(I_n-KH)\bigr)^{-1} c,
    \label{eq:linear_bstar}
\end{equation}
which reduces to \eqref{eq:fixedpoint} when $B = 0$. This is the steady-state
mean error of a Kalman filter with a mismatched model
\citep[][eq.~(29)]{lee_discrete-time_1979}.

\subsubsection{Nonlinear bias}

For a general nonlinear $b$, closing \eqref{eq:sd_exact} requires an
approximation. Since the true state has zero mean at stationarity
(Section~\ref{sec:statedep_setup}), the stationary mean of the analysis is
$\bar{x}^\mathrm{a} := \lim_{i\to\infty}\mathbb{E}[x^\mathrm{a}_i] = (I_n-KH)b^*$.

The model bias depends on $x^\mathrm{a}_i$, which varies across realisations (equivalently, at stationarity, over a long time series). Unless
$b$ is affine (Section~\ref{sec:linear_sd}), the average of $b$
over these varying analyses, $\mathbb{E}[b(x^\mathrm{a}_i)]$, differs from
its value at the average analysis, $b(\bar{x}^\mathrm{a})$. Ignoring this variation,
that is, replacing $\mathbb{E}[b(x^\mathrm{a}_i)]$ by $b(\bar{x}^\mathrm{a})$, gives the fixed-point equation
\begin{equation}
    b^* = (I_n-\Phi)^{-1} b\bigl((I_n-KH)b^*\bigr).
    \label{eq:nl_fp}
\end{equation}
We call this the zeroth-order approximation, because it keeps only the leading term of a Taylor expansion of $b$ about $\bar{x}^\mathrm{a}$.

Accounting for the variation of the analysis requires its stationary covariance,
$\Sigma^\mathrm{a} := \lim_{i\to\infty}\mathrm{Cov}(x^\mathrm{a}_i)$, which differs
from $P^\mathrm{a}$. We approximate $\Sigma^\mathrm{a}$ of the filter with a biased model
by the covariance of the analysis of the filter with a perfect model. Denoting the perfect-model analysis by $x^{\mathrm{a},0}_i$, the analysis error
$x^{\mathrm{a},0}_i - x^\mathrm{t}_i$ has covariance $P^\mathrm{a}$ asymptotically and
is uncorrelated with $x^{\mathrm{a},0}_i$
\citep[][Section~5.2, Theorem~2.5]{anderson_optimal_1979}, so that
$\lim_{i\to\infty}\mathrm{Cov}(x^{\mathrm{a},0}_i) = \Sigma^\mathrm{t} - P^\mathrm{a}$. Here $\Sigma^\mathrm{t}$ is the stationary
covariance of the true state, which exists by the ergodicity assumption of
Section~\ref{sec:statedep_setup} and solves $\Sigma^\mathrm{t} = M\Sigma^\mathrm{t}M^T + Q$. For the biased model, the approximation
$\Sigma^\mathrm{a} \approx \Sigma^\mathrm{t} - P^\mathrm{a}$ is exact for a constant
bias, which only shifts the analysis; a state-dependent bias adds a random component
to the forecast error, which changes $\Sigma^\mathrm{a}$. Unlike $P^\mathrm{a}$, $\Sigma^\mathrm{a}$ does not shrink as
the observations improve: it approaches $\Sigma^\mathrm{t}$, the natural variability
of the system. When this is not small relative to the scale on which $b$ varies, the
zeroth-order approximation is poor.

The analysis of the filter with a perfect model, $x^{\mathrm{a},0}_i$, is Gaussian, as a linear function of Gaussian variables. If the bias does not strongly distort this distribution, we can approximate $x^\mathrm{a}_i \sim \mathcal{N}(\bar{x}^\mathrm{a}, \Sigma^\mathrm{a})$ for large $i$ in \eqref{eq:sd_exact}, giving the Gaussian closure
\begin{equation}
    b^* = (I_n-\Phi)^{-1}\,
    \mathbb{E}_{X\sim\mathcal{N}((I_n-KH)b^*,\,\Sigma^\mathrm{a})}\bigl[b(X)\bigr].
    \label{eq:nl_fp_gauss}
\end{equation}

\subsubsection{Correction with the averaged increment}\label{sec:avg_correction}

Correcting the model with the averaged increment, as in Section~\ref{sec:bias_correction} for a constant bias, can fail when the bias is state-dependent. Even for a linear bias, Propositions~\ref{prop:mahal}
and~\ref{prop:iterated_corr} need not hold: subtracting a constant
correction from $c$ gives a corrected forecast bias governed by
$\bigl(I_n - \Phi - B(I_n-KH)\bigr)^{-1}$ rather than $(I_n-\Phi)^{-1}$, to which the
identities of Lemma~\ref{lem:key_identity} do not apply. This means that a constant correction can increase the innovation bias in this setting, and the iterated correction can diverge. Consistent with this, in experiments with a quasi-geostrophic model, \citet{delsole_empirical_1999} found that a constant correction did not improve forecast skill when the model error was mainly state-dependent.

\subsection{Decomposition of the increments}\label{sec:pathwise}

Averaging increments over the whole record precludes state-dependent corrections. To understand the state-dependent information contained in the increments, we need a decomposition of each increment, not only its mean. Consider a
second filter with the same gain and the
same observations, but with a perfect forecast model,
$x^{\mathrm{f},0}_{i+1} = Mx^{\mathrm{a},0}_i$. Let
$\Delta_i := x^\mathrm{f}_i - x^{\mathrm{f},0}_i$ be the difference between the
two forecasts, that is, the contribution of the bias to the forecast error.

\begin{proposition}[Decomposition of the increments]\label{prop:pathwise}
    Under the assumptions of Section~\ref{sec:setup}, for any $b(\cdot)$,
    \begin{subequations}
    \begin{equation}
        \Delta_{i+1} = \Phi\Delta_i + b(x^\mathrm{a}_i),
        \qquad\text{so that}\qquad
        \Delta_i = \Phi^i\Delta_0 + \sum_{k=0}^{i-1}\Phi^k\, b(x^\mathrm{a}_{i-1-k}),
        \label{eq:pathwise}
    \end{equation}
    and each increment can be decomposed as
    \begin{equation}
        x^\mathrm{a}_i - x^\mathrm{f}_i = -KH\Delta_i + \nu_i,
        \qquad \nu_i := x^{\mathrm{a},0}_i - x^{\mathrm{f},0}_i .
        \label{eq:pathwise_inc}
    \end{equation}
    \end{subequations}
    Here $\nu_i$ generalises the fluctuation of the same name in
    Section~\ref{sec:increment}.
\end{proposition}
\begin{proof}
    Both filters apply the analysis update
    $x^\mathrm{a} = (I_n-KH)x^\mathrm{f} + Ky$ to the same observations, so
    $x^\mathrm{a}_i - x^{\mathrm{a},0}_i = (I_n-KH)\Delta_i$, and the forecast
    step gives $\Delta_{i+1} = M(I_n-KH)\Delta_i + b(x^\mathrm{a}_i)$; unrolling
    gives \eqref{eq:pathwise}. The increment of the biased filter is
    $K(y_i - Hx^\mathrm{f}_i) = K(y_i - Hx^{\mathrm{f},0}_i) - KH\Delta_i$, which is
    \eqref{eq:pathwise_inc}.
\end{proof}

The decomposition involves no expectation. Its second term, $\nu_i$, is the increment
of the filter with a perfect model, that is, $K$ times its innovation.
These innovations are asymptotically distributed as $\mathcal{N}(0, S)$ and
independent of all earlier observations
\citep[][Section~5.3, Theorem~3.1]{anderson_optimal_1979}, and $x^\mathrm{f}_i$ is
a function of those observations. Hence $\nu_i$ is asymptotically distributed as
$\mathcal{N}(0, KSK^T)$, as in Section~\ref{sec:increment}, and independent of
$x^\mathrm{f}_i$. The first term, $-KH\Delta_i$, is the model error along the recent
trajectory, weighted by $\Phi^k$, and projected by $KH$.
For constant $b$, $\Delta_i$ is linear in $b$, and \eqref{eq:pathwise_inc} is the
decomposition used by \citet[][Section~III]{friedland_treatment_1969} to compute
the optimal estimate of a constant bias from the innovations of a filter that
ignores it; it then reduces to the split \eqref{eq:increment_decomposition}, with
$\Delta_i = b^\mathrm{f}_i$ asymptotically.

\subsection{Increments averaged by slow state}\label{sec:slow}

Between averaging the increments over the whole record and the individual increments lies an
intermediate case. Suppose that the model error depends on the state mainly through
components that vary slowly, such as those carrying a seasonal cycle or a regime that
persists for many assimilation cycles, and denote these slow components by $s$. By \eqref{eq:pathwise}, the
contribution of the bias to the forecast error, $\Delta_i$, is a weighted sum of past
values of the model error along the trajectory, with weights $\Phi^k$ that decay over
about $\tau := 1/(1-\rho(\Phi))$ cycles. If $s$ changes little over $\tau$ cycles, that
is, if it varies on a timescale of $L \gg \tau$ cycles, we can make a quasi-static approximation by noting that $\Delta_i$ stays close, on average, to
$(I_n-\Phi)^{-1}\mathbb{E}[b(x^\mathrm{a}_{i-1}) \mid s_i]$, the value it would settle at
if $s$ were held fixed at $s_i$. Here $s_i := s(x^\mathrm{f}_i)$ denotes the slow
components of the forecast.

Averaging the increments over cycles with similar values of $s_i$ then gives
\begin{equation}
    \mathbb{E}[x^\mathrm{a}_i - x^\mathrm{f}_i \mid s_i]
    \approx -KH(I_n-\Phi)^{-1}\,\mathbb{E}[b(x^\mathrm{a}_{i-1}) \mid s_i],
    \label{eq:quasistatic}
\end{equation}
so that the averaged increment, as a function of $s_i$, diagnoses the slowly varying
part of the model error.

An adaptive scheme such as that of \citet{dee_bias_2005} (Section~\ref{sec:dee})
can also leverage this idea: with small $\gamma$, it averages the increments
over roughly the last $1/\gamma$ cycles, and so tracks the part of the model error
that changes slowly compared with this window and with the memory $\tau$ of the
filter, but not its dependence on the fast-changing state.

\subsection{Individual increments}\label{sec:nl_bound}

We now move beyond the slow-state averaging to learning the full state-dependent correction. The natural target is the part of the increment that can
be predicted from the forecast state, the conditional mean
$\mathbb{E}[x^\mathrm{a}_i - x^\mathrm{f}_i \mid x^\mathrm{f}_i]$. The machine
learning approaches cited in the introduction estimate it by least-squares
regression: given a record of $T$ forecasts and analyses, they fit a correction
function $\hat{f} : \mathbb{R}^n \to \mathbb{R}^n$ over a class $\mathcal{F}$ of
functions, such as neural networks,
\begin{equation*}
    \hat{f} = \operatorname*{arg\,min}_{f \in \mathcal{F}} \frac{1}{T}\sum_{i=1}^{T}
    \bigl\|(x^\mathrm{a}_i - x^\mathrm{f}_i) - f(x^\mathrm{f}_i)\bigr\|_2^2,
\end{equation*}
which approaches the conditional mean for a long record and a sufficiently flexible
class. This follows from the ergodicity assumption of Section~\ref{sec:statedep_setup} and the classical result that the conditional mean minimises the expected squared error \citep[][Section~2.3, Theorem~3.2]{anderson_optimal_1979}. Once learned, the predicted increment
$\hat{f}(x^\mathrm{f}_i)$ is added to the forecast $x^\mathrm{f}_i$. We now ask what this conditional
mean, and each increment, represent, first for linear and then for nonlinear
dynamics.

\subsubsection{Linear dynamics}

Taking the conditional mean of \eqref{eq:pathwise_inc} gives
$\mathbb{E}[x^\mathrm{a}_i - x^\mathrm{f}_i \mid x^\mathrm{f}_i] = -KH\,\mathbb{E}[\Delta_i \mid x^\mathrm{f}_i]$, since $\nu_i$ has zero mean and is
independent of $x^\mathrm{f}_i$ (Section~\ref{sec:pathwise}). Each increment is therefore an unbiased but noisy measurement of
$-KH\Delta_i$. If $b$ changes little along the trajectory
over the memory time $\tau$, then
$\Delta_i \approx (I_n-\Phi)^{-1}b(x^\mathrm{a}_{i-1})$, and each increment
approximates $-KH(I_n-\Phi)^{-1}b(x^\mathrm{a}_{i-1})$. Such a regression therefore
recovers the model error as filtered by the assimilation: accumulated over the
memory of the filter and projected by $KH$, which, as in
Section~\ref{sec:increment}, cannot see the part of the accumulated bias in
$\ker(H)$.

\subsubsection{Nonlinear dynamics}

For nonlinear dynamics, the truth evolves as
$x^\mathrm{t}_{i+1} = \mathcal{M}(x^\mathrm{t}_i) + \eta_i$, the forecast is
$x^\mathrm{f}_{i+1} = \mathcal{M}(x^\mathrm{a}_i) + b(x^\mathrm{a}_i)$, and the filter
with a perfect model uses
$x^{\mathrm{f},0}_{i+1} = \mathcal{M}(x^{\mathrm{a},0}_i)$, so that
$\mathcal{M}(x) = Mx$ recovers the setting above. With a fixed gain, the split
\eqref{eq:pathwise_inc} still holds exactly, since the analysis step is linear, and
taking its conditional mean gives
\begin{equation}\label{eq:conditional_split}
    \mathbb{E}[x^\mathrm{a}_i - x^\mathrm{f}_i \mid x^\mathrm{f}_i]
    = -KH\,\mathbb{E}[\Delta_i \mid x^\mathrm{f}_i]
    + \mathbb{E}[\nu_i \mid x^\mathrm{f}_i].
\end{equation}
The first term is the contribution of the model error, as in the linear case. The
contribution of the bias to the forecast error now satisfies
$\Delta_{i+1} = b(x^\mathrm{a}_i) + \mathcal{M}(x^\mathrm{a}_i) - \mathcal{M}(x^{\mathrm{a},0}_i)$:
the model error of the latest step plus that of earlier steps, propagated by the
model, in place of the explicit form \eqref{eq:pathwise}. As before, the model error is seen only through $KH$, although for
nonlinear dynamics bias in the unobserved directions need not be invisible
(Remark~\ref{rem:nonlinear_ident}).
The second term is the conditional mean of $\nu_i = x^{\mathrm{a},0}_i - x^{\mathrm{f},0}_i$, and it represents the
mean increment that the filter would make with a perfect model, given the forecast.
For linear dynamics with Gaussian noise it vanishes, but in general it does not. A regression
on the increments learns this term along with the model error.

\subsubsection{Iterating the learned correction}

As for a constant bias in Section~\ref{sec:iterated}, the learned correction
can be iterated. After the correction is learned, it is applied to the forecast model. The filter is then
rerun with the corrected model, and the correction is retrained on the new increments plus the current correction. This can start from the previous iteration, so that the network is fine-tuned rather than retrained from scratch. Most applications, such as
those cited in the introduction and \citet{brajard_combining_2021}, stop after one
step; \citet{gregory_machine_2024} iterate three times.

If the iteration converges,
it follows from Proposition~\ref{prop:pathwise} that, for linear dynamics, it stops
when the remaining model error, accumulated over the memory of the filter, is on
average invisible to the observations for each forecast state. This is the analogue
of the condition $Hb^\mathrm{f} = 0$ for a constant bias and, as there, does not
imply that the remaining model error is zero. Whether the iteration converges does not appear to have been studied.

\section{Numerical experiments}\label{sec:experiments}
\subsection{Linear test system}
\label{sec:linear_system}

The experiments of Sections~\ref{sec:avg_exp}, \ref{sec:iter_exp}, and~\ref{sec:statedep_exp} use a
Kalman filter (KF) operating at steady state, and those of Section~\ref{sec:avg_exp} also a
stochastic ensemble Kalman filter (EnKF) with perturbed observations
\citep[][Section~3.4]{evensen_ensemble_2003}, on a linear system. The state dimension is $n = 4$ and the observation dimension is $p = 2$. The
model $M$ consists of two damped rotations, with $\rho(M) = 0.95$, and each
observation is the sum of two state components, so that $\ker(H)$ is two-dimensional. The observation and model error covariances, $R$ and $Q$, are multiples of the
identity. The parameter values, including those of the biases below, are listed in
Table~\ref{tab:settings} (Appendix~\ref{appendix:settings}).

The steady-state KF gain is computed from the discrete algebraic Riccati
equation \eqref{eq:dare}. This gives
$\rho(\Phi) = 0.805$.
The EnKF uses multiplicative covariance inflation, and model noise
$\eta_i \sim \mathcal{N}(0, Q)$ is drawn independently for each member's forecast.

\subsection{Constant bias}
\label{sec:const_exp}

\subsubsection{Averaged increment}
\label{sec:avg_exp}

\begin{figure}
    \centering
    \includegraphics[width=\linewidth]{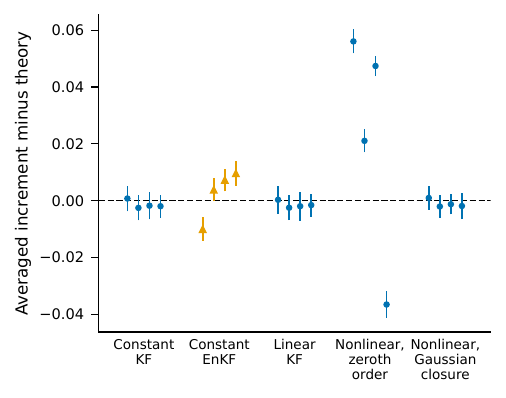}
    \caption{Time-averaged analysis increment minus its theoretical value $-KHb^*$
    from \eqref{eq:increment}, for each of the four state components (the four points in
    each group). Here $b^*$ is given by \eqref{eq:fixedpoint} for the constant bias
    of Section~\ref{sec:avg_exp} (KF and EnKF), by \eqref{eq:linear_bstar}
    for the linear bias, and by the zeroth-order approximation \eqref{eq:nl_fp} or
    the Gaussian closure \eqref{eq:nl_fp_gauss} for the nonlinear bias of
    Section~\ref{sec:statedep_exp} (KF). Error bars show $\pm 1$ standard error.}
    \label{fig:validation}
\end{figure}

We verify the fixed-point expressions \eqref{eq:fixedpoint} and
\eqref{eq:increment} numerically for both filters on the linear system of
Section~\ref{sec:linear_system}, with the bias $b$ given in Table~\ref{tab:settings}.
Figure~\ref{fig:validation} compares the time-averaged analysis increment
with theory for each filter. For the KF, the time-averaged forecast error
agrees with the asymptotic forecast bias $b^*$, and the averaged increment with
$-KHb^*$, to within one standard error in every component. The standard error is
estimated from the spread of the means over equal, non-overlapping segments of the
record, each much longer than the memory of the filter, which accounts for the
correlation between successive cycles. The averaged EnKF increment also agrees with $-KHb^*$, but
differs slightly more because of
sampling error with a finite ensemble.

\subsubsection{Iterated correction}
\label{sec:iter_exp}

\begin{figure}
    \centering
    \includegraphics[width=\linewidth]{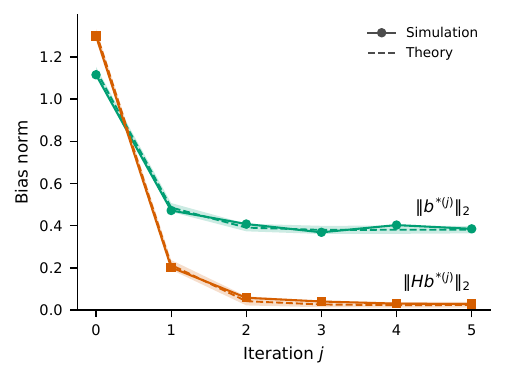}
    \caption{Convergence of the iterated analysis increment correction. Green circles: full-state bias norm $\|b^{*(j)}\|_2$.
    Orange squares: innovation bias norm $\|Hb^{*(j)}\|_2$. Markers show
    numerical results. Dashed lines show the median, and the shading the central 68\%
    range, of the norms
    predicted by \eqref{eq:iter_bias} and \eqref{eq:iter_innov},
    with the sampling noise of the finite-time averages included.}
    \label{fig:iteration}
\end{figure}

We verify the iterated correction convergence (Section~\ref{sec:iterated}) numerically,
with the KF on the same system. At each
iteration $j$, the filter is run with the
forecast model with per-step bias $b^{(j)}$, the bias estimate
$\hat{b}^{(j)} = -\overline{x^{\mathrm{a},(j)} - x^{\mathrm{f},(j)}}$ is computed from the
time-averaged analysis increment, and
subtracted from the per-step bias to give $b^{(j+1)}$. Figure~\ref{fig:iteration} shows both the full-state bias norm
$\|b^{*(j)}\|_2$ and the innovation bias norm $\|Hb^{*(j)}\|_2$
over five iterations, compared to the theoretical predictions
obtained by iterating \eqref{eq:iter_bias} and from \eqref{eq:iter_innov}.

The full state bias $\|b^{*(j)}\|_2$ drops
sharply on the first iteration then plateaus at $\approx 0.39$, in agreement
with the limit $\|\Pi b^*\|_2 = 0.38$ of Corollary~\ref{cor:iter_limit}. This
residual lies in $\ker(H)$, and it cannot be removed by any correction. By contrast, the innovation bias $\|Hb^{*(j)}\|_2$
converges to zero geometrically, as guaranteed by
Proposition~\ref{prop:iterated_corr}: it decreases by a factor of about $0.16$
in the first iteration, matching the rate $\rho(I_p - C) = 0.162$. From iteration $3$ onwards it remains at the level of
sampling noise from the finite averaging window.

\subsubsection{Nonlinear dynamics}
\label{sec:l96}

The theory of Section~\ref{sec:constant_bias} assumes linear dynamics and a fixed
gain. To check whether the iterated correction still works in this setting, we repeat the experiment of Section~\ref{sec:iter_exp} with the
Lorenz '96 model \citep{lorenz_predictability_1996},
\begin{equation*}
    \frac{dx_k}{dt} = (x_{k+1} - x_{k-2})x_{k-1} - x_k + F,
    \quad k = 1,\ldots,n,
\end{equation*}
with $n=20$, $F=8$, and periodic boundary conditions, integrated with a fourth-order Runge--Kutta scheme. Every
other variable is observed ($p = 10$), and a localised EnKF
\citep{houtekamer_sequential_2001} with multiplicative inflation and Gaspari--Cohn
localisation \citep{gaspari_construction_1999} is used. The forecast model has a
constant, spatially varying bias, a sinusoid in $k$ plus a small random part, whose
largest component, expressed as a tendency, is about $15\%$ of the forcing $F$. The
parameter values are listed in Table~\ref{tab:settings_l96}
(Appendix~\ref{appendix:settings}).

\begin{figure}
    \centering
    \includegraphics[width=\linewidth]{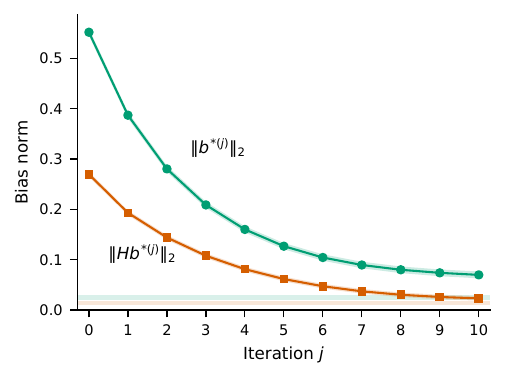}
    \caption{Iterated increment correction for the Lorenz '96 model with a constant,
    spatially varying bias. Green circles:
    full-state bias norm $\|b^{*(j)}\|_2$. Orange squares: innovation bias norm
    $\|Hb^{*(j)}\|_2$. The shading along the curves shows $\pm 1$ standard error, and the
    horizontal bands of the same colours the central $68\%$ range of each norm
    expected from sampling noise alone with a perfect model.}
    \label{fig:l96}
\end{figure}

Here $b^*$ is, as in Section~\ref{sec:averaged}, the expected forecast error at
stationarity, estimated by the time-averaged forecast error.
Figure~\ref{fig:l96} shows that the behaviour of Section~\ref{sec:iter_exp} carries
over. The innovation bias norm $\|Hb^*\|_2$ decreases geometrically, and approaches the level of sampling noise. The
full-state bias norm $\|b^*\|_2$ also decreases at first, but then levels off above the
perfect-model level. A residual bias, almost entirely in the unobserved components, thus remains.

\subsection{State-dependent bias}
\label{sec:statedep_bias_exp}

\subsubsection{Averaged increments}
\label{sec:statedep_exp}

We verify the fixed-point expressions of Section~\ref{sec:averaged} numerically
for both a linear and a nonlinear bias function, using the system of
Section~\ref{sec:linear_system}.

\paragraph{Linear bias.} We take $b(x) = Bx + c$, with $B$ and $c$ given in
Table~\ref{tab:settings}. The theoretical fixed point is given by
\eqref{eq:linear_bstar}.

\paragraph{Nonlinear bias.} We take $b(x) = c + \alpha\sin(x)$
elementwise. Then, in the forecast step, the filter applies \eqref{eq:sd_model},
but it keeps the steady-state Kalman gain of Section~\ref{sec:linear_system}.

Figure~\ref{fig:validation} compares the time-averaged analysis increment with
theory for both bias functions. For the linear bias, the theory is exact, and it
agrees with the averaged increment to within one standard error in every component.
For the nonlinear bias, the Gaussian closure also agrees to within one standard
error, whereas the zeroth-order approximation, which ignores the spread of the
analysis state, does not.

\subsubsection{Learned corrections}
\label{sec:learned_exp}

To illustrate Section~\ref{sec:nl_bound}, we learn a state-dependent correction
from the increments of the Lorenz '96 system of Section~\ref{sec:l96}. Here the bias depends on the state through the
same sinusoidal function in every component. The other settings are as in Section~\ref{sec:l96}, except for the bias, the
ensemble size, and the spin-up (Table~\ref{tab:settings_l96}). The correction is a neural network
whose input is the forecast value of a single component and whose output is the predicted
increment of that component, with two hidden layers. It is fitted by least squares, as in Section~\ref{sec:nl_bound}, to the increments of
the ensemble mean in the observed components, with the ensemble-mean forecast as
input, and is applied to those components only. In the filter, the network is evaluated at the
forecast of each ensemble member, and its output is added to that forecast.

To compare with the theory \eqref{eq:conditional_split}, we run a second filter with a perfect model and the same
gains and observations, which gives $-K_iH\Delta_i$ at every cycle. Its conditional mean given the forecast is estimated by fitting a network of the same form to $-K_iH\Delta_i$ in place of the increment.

\begin{figure*}
    \centering
    \includegraphics[width=0.9\linewidth]{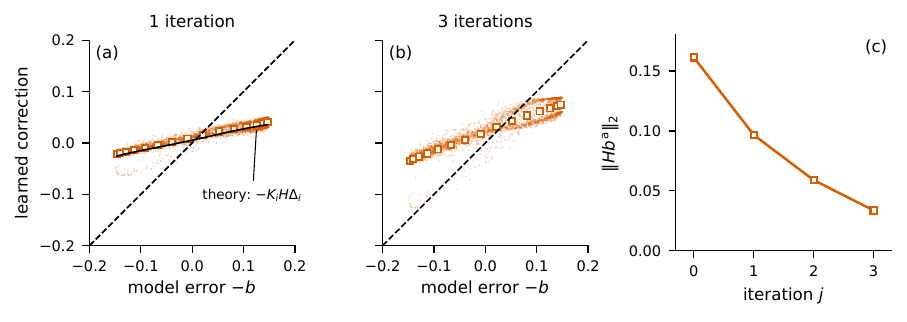}
    \caption{Learned corrections in the Lorenz '96 experiment with a state-dependent bias.
    (a) First learned correction against the model error $-b$ in the observed
    components, for individual cycles (points) and averaged in bins of $-b$ (squares).
    The solid line is the theory, the conditional mean of $-K_iH\Delta_i$ given the
    forecast, and the dashed line is $y = x$. (b) Same as (a), but after three iterations of
    the correction. (c) Norm of the analysis bias in the observed components, $\|Hb^\mathrm{a}\|_2$,
    estimated from the time-mean analysis error, against the number of iterations.
    The line shows the mean of two independent runs, which differ by at most $0.005$.}
    \label{fig:learned}
\end{figure*}

Figure~\ref{fig:learned}(a) shows the first learned correction in the observed
components. It follows the theory closely and demonstrates that, as per Section~\ref{sec:nl_bound}, the regression learns the model error
as filtered by the assimilation, not the model error itself. The small remaining
difference is the conditional mean of $\nu_i$, the systematic increment that the
filter would make with a perfect model. Iterating the correction as in Section~\ref{sec:nl_bound} brings it closer to $-b$ (Figure~\ref{fig:learned}(b)). The
analysis bias in the observed components falls about fivefold
(Figure~\ref{fig:learned}(c)).

\section{Conclusions}\label{sec:conclusions}

Analysis increments are a widely used and inexpensive way to diagnose and correct model bias. This paper has provided a theoretical foundation for their use.

For a linear model with a constant bias, and observations assimilated with a Kalman filter, we showed that the forecast bias settles to a steady-state value. The averaged increment provides an estimate of this bias, seeing the part of it that is visible in the
observations, spread into unobserved variables by the Kalman gain. Correcting
the model with the averaged increment reduces the bias but does not
remove it. Repeating the correction, each time using the increments of the
corrected model, removes the bias seen by the observations completely, at a
geometric rate. What remains is equivalent to a constant offset of the state in
directions that the observations do not see, and no method can
detect it. The repeated correction provides an alternative to online methods, such as augmented-state filters and likelihood-based methods, and does not require changing the assimilation system.

For state-dependent biases, which machine-learned corrections target,
the picture is similar but less complete. For linear dynamics, each increment
records the model error accumulated over the memory of the filter, plus noise. As for
a constant bias, it sees only the part visible in the observations, spread into
unobserved variables by the Kalman gain. Averaged over a long record, the increments measure the mean of this filtered error over the states the system visits. Unlike for a constant bias, correcting the model with this mean can increase the bias seen by the observations. Averaged separately for each season or weather regime, the increments recover
the slowly varying part of the model error, provided it changes slowly compared
with the memory of the filter. The individual increments can be used for learning state-dependent
corrections; however, they learn the filtered model error, not the model error
itself.

Numerical experiments confirm the theory for linear systems. Experiments with the Lorenz '96 model and an ensemble Kalman filter suggest that the main conclusions carry over to nonlinear systems.

Several questions remain open. Our results assume a filter that
is unaware of the bias and has a fixed gain; extending them to filters that adapt
their gain to the bias \citep{dee_data_1998} is an area for future work. Most of the analysis in this work focuses on linear systems, and extending it to the fully nonlinear
setting would bring the theory closer to operational systems and to
machine-learned corrections \citep{bonavita_machine_2020, gregory_deep_2023,
mojgani_interpretable_2024, bach_machine_2025, arcucci_convergence_2026}. For nonlinear dynamics, part of the bias that is undetectable for linear models can become detectable (Remark~\ref{rem:nonlinear_ident}). How much can be recovered is an open question. Statistics beyond the mean, such as lagged innovation covariances \citep{bonavita_exploring_2021}, or learned corrections with memory, which use the recent history of the forecasts rather than only the current one \citep{levine_framework_2022}, may help.
A similar analysis could be applied to increments from smoothers such as 4D-Var or the Kalman smoother. Finally, the analysis could be extended to observations of statistics of the
state rather than the state itself \citep{bach_filtering_2024} and to multi-model
data assimilation \citep{bach_multi-model_2023}.

\section*{Acknowledgements}

I am grateful to the late Eugenia Kalnay, whose work with her students and collaborators on estimating and correcting model error from analysis increments motivated this work.

\section*{Declaration of Generative AI and AI-assisted technologies in the writing process}

During the preparation of this work, the author used Claude Sonnet 4.6, Opus 5, and Opus 5.5 (Anthropic) to
assist with mathematical exposition and proof development, as well as
implementation of numerical experiments. After using these tools, the author reviewed and edited all content,
verified all mathematical arguments and numerical results, and takes full responsibility
for the content of the publication.

\bibliography{references}

\appendix

\section{Proofs}\label{appendix:proof}

Throughout, $S = HP^\mathrm{f}H^T + R$, $K = P^\mathrm{f}H^TS^{-1}$, $\Phi = M(I_n - KH)$, and $C = H(I_n-\Phi)^{-1}K$. Recall that
$\rho(\Phi) < 1$, so that $X := (I_n - \Phi)^{-1}$ exists, and that $P^\mathrm{f}$ and $R$ are positive definite. For symmetric matrices, $A \preceq B$ means that $B - A$ is positive semidefinite. Proposition~\ref{prop:mahal} uses the following identities, in which $W := I_p - HXMK$.

\begin{lemma}[Key identities]\label{lem:key_identity}
    \begin{equation}
        I_p - C = WRS^{-1}, \qquad S = WRW^T + HXQX^TH^T.
        \label{eq:key_id}
    \end{equation}
\end{lemma}
\begin{proof}
Since $KS = P^\mathrm{f}H^T$, we have $HK = (S - R)S^{-1}$, that is,
$RS^{-1} = I_p - HK$. Hence
\begin{equation*}
    WRS^{-1} = (I_p - HXMK)(I_p - HK) = I_p - HK - HXMK(I_p - HK).
\end{equation*}
Since $MK(I_p - HK) = M(K - KHK) = M(I_n - KH)K = \Phi K$, this equals
$I_p - H(I_n + X\Phi)K = I_p - HXK = I_p - C$, using $I_n + X\Phi = X$. The second
identity is
\citet[][Section~4.5, Problem~5.1]{anderson_optimal_1979} at $z = 1$.
\end{proof}

\begin{proof}[Proof of Proposition~\ref{prop:mahal}]
From \eqref{eq:corrected_bias},
$H\tilde{b}^* = H\bigl(I_n-(I_n-\Phi)^{-1}KH\bigr)b^* = (I_p-C)Hb^*$. By the second
identity in \eqref{eq:key_id}, $WRW^T \preceq S$, so $Y := S^{-1/2}WR^{1/2}$
satisfies $YY^T \preceq I_p$, and hence $Y^TY \preceq I_p$, that is,
$R^{1/2}W^TS^{-1}WR^{1/2} \preceq I_p$. By the first identity,
\begin{align*}
    (I_p-C)^TS^{-1}(I_p-C)
    &= S^{-1}R^{1/2}\bigl(R^{1/2}W^TS^{-1}WR^{1/2}\bigr)R^{1/2}S^{-1}\\
    &\preceq S^{-1}RS^{-1}.
\end{align*}
Moreover,
$S^{-1} - S^{-1}RS^{-1} = S^{-1}(S - R)S^{-1} = S^{-1}HP^\mathrm{f}H^TS^{-1} \succeq 0$. If $v \neq 0$ is in $\mathrm{range}(H)$, so that $v = Hz$ for some $z \in \mathbb{R}^n$,
then $u := H^TS^{-1}v \neq 0$, since $z^Tu = (Hz)^TS^{-1}v = v^TS^{-1}v > 0$. Then,
$v^T(S^{-1} - S^{-1}RS^{-1})v = u^TP^\mathrm{f}u > 0$. Hence, for every nonzero
$v \in \mathrm{range}(H)$,
\begin{equation*}
    \|(I_p-C)v\|_S^2 \leq v^TS^{-1}RS^{-1}v = \|v\|_S^2 - u^TP^\mathrm{f}u < \|v\|_S^2,
\end{equation*}
that is,
\begin{equation}
    \|(I_p-C)v\|_S < \|v\|_S \quad \text{for every nonzero } v \in \mathrm{range}(H).
    \label{eq:strict_contraction}
\end{equation}
Taking $v = Hb^*$ gives the result, with strict inequality whenever
$Hb^* \neq 0$.
\end{proof}

\begin{proof}[Proof of Proposition~\ref{prop:iterated_corr}]
By \eqref{eq:strict_contraction}, $I_p - C$ shrinks the $S$-norm of every nonzero
vector in $\mathrm{range}(H)$, which it maps into itself, since
$(I_p - C)Hz = H(z - XKHz)$. Since $\mathrm{range}(H)$ is finite-dimensional,
$I_p - C$ shrinks all such vectors by at least a common factor $\vartheta < 1$:
$\|(I_p-C)v\|_S \leq \vartheta\|v\|_S$ for all $v \in \mathrm{range}(H)$. Hence
$(I_p-C)^jv \to 0$ geometrically for every $v \in \mathrm{range}(H)$, and in
particular $\|Hb^{*(j)}\|_S$ decreases monotonically to zero. If $H$ has full row
rank, $\mathrm{range}(H) = \mathbb{R}^p$, so
$(I_p-C)^TS^{-1}(I_p-C) \preceq \vartheta^2 S^{-1}$. For an eigenvalue $\mu$ of
$I_p - C$ with eigenvector $x \in \mathbb{C}^p$,
\begin{equation*}
    |\mu|^2\, x^*S^{-1}x = x^*(I_p-C)^TS^{-1}(I_p-C)\,x \leq \vartheta^2\, x^*S^{-1}x,
\end{equation*}
so $|\mu| \leq \vartheta$, and $\rho(I_p - C) \leq \vartheta < 1$.
\end{proof}

\begin{proof}[Proof of Corollary~\ref{cor:iter_limit}]
    Each iteration subtracts $KHb^{*(j)} = K(I_p-C)^jHb^*$ from the per-step
    bias, by \eqref{eq:iter_innov}. Since $Hb^* \in \mathrm{range}(H)$, on which
    $(I_p-C)^j \to 0$ geometrically (Proposition~\ref{prop:iterated_corr}), the
    Neumann series gives $\sum_{j\geq0}(I_p-C)^jHb^* = C^{-1}Hb^*$, so
    $b^{(j)}\to b^{(\infty)}$ and $b^{*(j)} = (I_n-\Phi)^{-1}b^{(j)} \to \Pi b^*$.
    Since $H(I_n-\Phi)^{-1}KC^{-1}H = CC^{-1}H = H$, we have $H\Pi = 0$. Hence $z := \Pi b^*$ satisfies $Hz = 0$, and
    $b^{(\infty)} = (I_n-\Phi)z = (I_n-M)z$, since $\Phi z = Mz$ when $Hz = 0$.
\end{proof}

\section{Experimental settings}\label{appendix:settings}

Table~\ref{tab:settings} lists the parameters of the
experiments with the linear system
(Sections~\ref{sec:avg_exp}, \ref{sec:iter_exp}, and~\ref{sec:statedep_exp}), and Table~\ref{tab:settings_l96} lists the parameters of the experiments with the Lorenz '96
model (Sections~\ref{sec:l96} and~\ref{sec:learned_exp}).

\begin{table}[!t]
\centering
\caption{Parameters of the experiments with the linear system. $\mathcal{R}(\theta)$
is the $2\times2$ rotation matrix by angle $\theta$, and $P$ the cyclic shift,
$(Px)_k = x_{k+1}$ with $x_5 := x_1$.}
\label{tab:settings}
\small
\begin{tabular}{@{}>{\raggedright\arraybackslash}p{0.25\columnwidth}>{\raggedright\arraybackslash}p{0.69\columnwidth}@{}}
\toprule
Model & $M = 0.95\,\mathrm{diag}\bigl(\mathcal{R}(\theta), \mathcal{R}(2\theta)\bigr)$, $\theta = 0.4$ rad \\
Observations & $H = (I_2\ \ I_2)$; $Q = 0.5\,I_4$, $R = 0.5\,I_2$ \\
Constant bias (Sections~\ref{sec:avg_exp} and~\ref{sec:iter_exp}) & $b = (0.5,\,-0.3,\,0.2,\,-0.4)^T$ \\
Linear bias (Section~\ref{sec:statedep_exp}) & $b(x) = Bx + c$, $B = 0.05\,(I_4 - P)$, $c = (0.3,\,-0.2,\,0.1,\,-0.15)^T$ \\
Nonlinear bias (Section~\ref{sec:statedep_exp}) & $b(x) = c + \alpha\sin(x)$ (elementwise), $\alpha = 0.3$ \\
EnKF (Section~\ref{sec:avg_exp}) & stochastic, $N = 50$, anomaly inflation $1.02$ \\
Run lengths & Sections~\ref{sec:avg_exp} and~\ref{sec:statedep_exp}: $T = 30\,000$, spin-up $500$; Section~\ref{sec:iter_exp}: $8\,000$ cycles per iteration, spin-up $200$ \\
\bottomrule
\end{tabular}
\end{table}

\begin{table}[!t]
\centering
\caption{Parameters of the Lorenz '96 experiments.}
\label{tab:settings_l96}
\small
\begin{tabular}{@{}>{\raggedright\arraybackslash}p{0.25\columnwidth}>{\raggedright\arraybackslash}p{0.69\columnwidth}@{}}
\toprule
Model & $n = 20$, $F = 8$; RK4 with $\Delta t = 0.05$  \\
Observations & $x_1, x_3, \ldots, x_{19}$ ($p = 10$); $Q = 0.01\,I_{20}$, $R = 0.5\,I_{10}$ \\
Constant bias (Section~\ref{sec:l96}) & $b_k = 0.04\sin\bigl(2\pi(k-1)/n\bigr) + 0.016\,\xi_k$, with $\xi_k \sim \mathcal{N}(0,1)$ drawn once \\
State-dependent bias (Section~\ref{sec:learned_exp}) & $b(x) = 0.15\sin(0.6x + 0.5)$ (elementwise) \\
EnKF & stochastic, covariance inflation $1.04$, Gaspari--Cohn localisation of half-width $5$;

Section~\ref{sec:l96}: $N = 100$;

Section~\ref{sec:learned_exp}: $N = 40$ \\
Run lengths & $T = 30\,000$ cycles;

Section~\ref{sec:l96}: spin-up $1\,000$;

Section~\ref{sec:learned_exp}: spin-up $2\,000$ \\
Network (Section~\ref{sec:learned_exp}) & one input, standardised, two hidden layers of $32$ $\tanh$ units, one output; Adam with learning rate $3\times10^{-3}$ ($10^{-3}$ when fine-tuning) and cosine decay, $40$ epochs, batches of $4\,096$; output layer refitted by least squares \\
\bottomrule
\end{tabular}
\end{table}

\end{document}